\documentclass[]{aa}
\usepackage[dvips]{graphicx}
\usepackage[]{txfonts}
\def\simless{\mathbin{\lower 3pt\hbox
{$\rlap{\raise 5pt\hbox{$\char'074$}}\mathchar"7218$}}}   
\def\simmore{\mathbin{\lower 3pt\hbox
{$\rlap{\raise 5pt\hbox{$\char'076$}}\mathchar"7218$}}}   
\newcommand{\be}{\begin{equation}}
\newcommand{\ee}{\end{equation}}

\begin{document}
\title{On the existence of a reverse shock in magnetized GRB ejecta}
\titlerunning{On the existence of reverse shock in magnetized GRB ejecta}
\authorrunning{}
\author{D. Giannios\inst{1}
\and P. Mimica\inst{2}
\and M.~A. Aloy\inst{2}}

\institute{Max Planck Institute for Astrophysics, Box 1317, D-85741 Garching,
  Germany
\and Departamento de Astronom\'ia y Astrof\'isica, Universidad de Valencia, 46100,
  Burjassot, Spain}

\offprints{giannios@mpa-garching.mpg.de}
\date{Received / Accepted}

\abstract{The role of magnetic fields in gamma-ray burst (GRB) flows
  remains controversial. The study of the early afterglow phases and,
  in particular, of the reverse shock dynamics and associated emission
  offers a promising probe of the magnetization of the ejecta. In this
  paper, we derive the conditions for the existence of a reverse shock
  in arbitrarily magnetized ejecta that decelerate and interact with the
  circumburst medium. Both constant and wind-like density profiles
  are considered. We show, in contrast to
  previous estimates, that ejecta with magnetization $\sigma_0\simmore
  1$ are not crossed by a reverse shock for a large fraction of the
  parameter space relevant to GRB flows.  Allowing for shell
  spreading, there is always a relativistic or mildly relativistic
  reverse shock forming in $\sigma_0\simless 0.3$ ejecta. From this,
  we conclude that the paucity of optical flashes, believed to be
  a distinctive signature of a reverse shock, may be 
  explained by the existence of dynamically important magnetic
  fields in the ejecta.
  \\

  }

\maketitle

\section{Introduction} 
\label{intro}

The role of magnetic fields in the GRB flow remains uncertain.
Neutrino annihilation may be the dominant process that leads to the
launching of the GRB flow by forming a fireball, i.e.  a flow
initially dominated by its thermal energy (Goodman 1986; Paczynski
1986). Alternatively, if they are powerful
enough and develop the appropriate topology, magnetic fields
can efficiently extract the rotational energy from the central engine launching a
Poynting-flux-dominated flow (hereafter PDF; Usov 1992).

In the fireball model the magnetic fields are dynamically
unimportant. The GRB flow during the onset of the afterglow phase is
expected to be, at most, weakly magnetized.  Though model dependent,
the conversion of magnetic into kinetic  energy in a PDF is partial
(Michel 1969; Li et al. 1992; Vlahakis \& K\"onigl 2003; Giannios \&
Spruit 2006). The flow is expected to have a significant
fraction of its energy in the form of Poynting flux at large radii
where there is substantial interaction with the external medium.

The study of the early phases of the ejecta interaction with
the external medium can probe the strength of magnetic fields in
flow. The early emission from the forward shock, for example, depends
on the magnetization of the ejecta (Lyutikov \& Blandford 2003; Zhang
\& Kobayashi 2005, hereafter ZK05; Genet et al. 2006).  Modeling of the microphysical
parameters of the forward and the reverse shock supports the existence of 
strong magnetic fields in the ejecta (Fan et al. 2002;
Kumar \& Panaitescu 2003; Zhang et al. 2003). Furthermore, polarization 
measurements can provide information on the structure of the magnetic fields
(Lazzati et al. 2003; Granot \& Taylor 2005; Mundell et al. 2007).
 
The early optical afterglow lightcurves depend critically on the
very existence of a reverse shock into strongly magnetized ejecta.
A reverse shock may also contribute in a dominant way 
to the observed emission in later afterglow phases 
(Genet et al. 2007; Uhm \& Beloborodov 2007).     
In this paper, we investigate the parameter space of the magnetization
of the flow for which there is a reverse shock crossing the ejecta. 
This parameter space is described by simple analytic expressions. 
We show that, in contrast to previous claims, if the flow contains similar
amounts of kinetic and magnetic energy at the onset of the afterglow, there
is no reverse shock forming in a large fraction of GRBs.
The observational implications of our findings are also discussed.

\section{The early afterglow phases}

\subsection{Magnetization of the flow at the onset of the afterglow}
\label{sec:magnetization}

For the GRB flow to be accelerated to ultrarelativistic speeds, 
it must be launched with a high energy-to-rest-mass ratio.
The primary form of the energy of the outflow remains unknown.  
In the fireball model (Goodman 1986; Paczynski 1986), most of 
the energy is assumed to be initially stored in thermal form while 
in PDFs (Usov 1992; M\'esz\'aros \& Rees 1997b; Spruit et al. 2001;
Van Puten \& Ostriker 2001; Uzdensky \& MacFadyen 2006) 
in magnetic energy density. 

Neutrino-antineutrino annihilation taking place in the polar region of
a hyper-accreting central object can lead to the formation of a
fireball (e.g., Woosley 1993; Ruffert \& Janka 1998; Aloy et al. 2000;
Aloy, Janka \& M\"uller 2005). The fireball is accelerated by its
internal pressure gradients that convert its thermal energy into
kinetic energy of the baryons. After the acceleration phase is over,
the flow may consist of faster and slower shells that collide with
each other. These collisions lead to internal shocks that power the
prompt emission (Paczynski \& Xu 1994; Rees \& M\'esz\'aros 1994;
Daigne \& Mochkovitch 1998; Mimica et al. 2005; Mimica, Aloy \& 
M\"uller 2007). Internal shocks can lead to
amplification of seed magnetic fields due to, e.g., the two-stream
instability (Medvedev \& Loeb 1999). The strength of these small
scale fields and whether their survival on macroscopic scales
downstream of the shocks is rather uncertain (e.g. Spitkovsky
2007). Still, they may account for $\sim 0.1-1$\% of the luminosity of
the flow. After the phase of internal collisions, the flow expands and
cools before it enters the afterglow phase.  There, the flow is
expected to be weakly magnetized (i.e. the magnetic energy density is
much less than the rest mass energy density of the flow).
 
On the other hand, magnetic fields can be efficient in extracting
rotational energy from the central engine which may be an accretion
disk (Blandford \& Payne 1982), a rotating black hole (Blandford \&
Znajek 1977) or a millisecond magnetar (Usov 1992).  The
resulting flow is initially Poynting-flux dominated. The acceleration
of a PDF is more model dependent than that of the fireball model. It
depends on the magnetic field geometry and on the role of magnetic
dissipation. In particular, magnetic dissipation can lead to efficient
acceleration of the flow by converting Poynting flux in to kinetic
energy (Drenkhahn \& Spruit 2002; Giannios \& Spruit 2006). Magnetic
dissipation can also efficiently power the prompt emission (Thompson
1994; Spruit et al. 2001; Lyutikov \& Blandford 2003; Giannios 2006b;
Thompson 2006; Giannios \& Spruit 2007). Despite the model dependent
aspects of MHD jet acceleration, a common outcome of different studies
is that the acceleration process is not expected to be 100\% efficient
in converting the Poynting flux into kinetic flux. As a result, the
flow remains strongly magnetized after the acceleration phase is over.
At large radii, magnetic fields carry energy comparable to that of the
kinetic energy of the baryons (Li et al. 1992); Drenkhahn \& Spruit
2002; Vlahakis \& K\"onigl 2003, Giannios \& Spruit 2006) or even
dominate over the kinetic energy of the flow (e.g. Lyutikov \&
Blandford 2003; Thompson 2006; Lyutikov 2006).

It has to be noted that the degree of magnetization of the ejecta may
critically depend on the likelihood of baryon mixing in the course of
the generation of a PDF, which can only be reliably determined by
extremely well resolved MHD numerical simulations. A substantial
baryon entrainment could result in a much smaller magnetization than
predicted by the above mentioned models. Furthermore, the
magnetization of PDFs could be substantially decreased by the
pair-loading resulting from $\nu\bar{\nu}$-annihilation close to the
central engine (see, e.g., Levinson \& Eichler 1993; Aloy \&
Obergaulinger 2007).

\subsection{Initial interaction of the ejecta with the external medium}

If far enough away from the central engine, there is
substantial interaction of the relativistic ejecta with the external
medium.  This interaction is believed to result to the afterglow
emission.  A characteristic difference between fireballs and PDFs lies
in the magnetization of the ejecta at the onset of the afterglow
phase. In fireball models, the energy of the flow is dominated by the
kinetic energy of baryons. If the flow is launched Poynting-flux dominated, it
is expected to maintain a large fraction of its energy in the form of
magnetic energy at large distance.

Since the initial phases of the ejecta-external medium interaction
depend on the magnetization of the ejecta (e.g., Lyutikov 2006), it is
possible distinguish among fireball and PDF models using early
afterglow observations.  A particularly promising probe of the
magnetic content of the flow comes from the study of the reverse shock
emission (e.g. Kumar \& Panaitescu 2003; Zhang et al. 2003). In this
work, we focus on the conditions for the existence of a reverse shock
in arbitrarily magnetized ejecta.

We focus on the interaction of the GRB ejecta with external medium
of constant number density  $n_{\rm e}$. Similar analysis can, however, 
be applied to different density profiles for the external medium.
The case of wind-like density profile $n\sim r^{-2}$ is also discussed in Sect. 3.2. 
For typical GRB parameters, this interaction becomes substantial at distances $r\sim
10^{15}-10^{17}$ cm. We assume that the acceleration and collimation of the flow has 
taken place much closer to the central engine as well as the internal dissipation 
mechanisms (e.g. internal shocks, magnetic dissipation) that are believed to power the prompt
emission. After internal dissipation is over, the flow expands and cools
down. Since we are interested in the afterglow phase, we consider a, {\it radially}
moving, {\it cold} and relativistic  shell with width $\Delta_0$  and 
corresponding Lorentz factor $\gamma_0\gg 1$. 

In a radially moving magnetized flow, the induction equation results
in a component of the magnetic field perpendicular to the motion
$B_{\perp} \propto 1/r$ that drops much slower with radius than
the parallel one: $B_{\parallel} \propto 1/r^2$. Since the flow
expands for several order of magnitude from the launching to the
afterglow region, we assume that the magnetic field is dominated by
its perpendicular component. The magnetic content of the shell is
conveniently parameterized by its magnetization $\sigma_0$ which
stands for the Poynting-to-kinetic flux ratio at some initial radius
$r_0$. For an ultrarelativistic flow \be \sigma_0\equiv
\frac{E_p}{E_k}=\frac{B_0^2}{4\pi \gamma_0\rho_0 c^2},
\label{sigma}
\ee 
 where $B_0$ and $\rho_0$ are the initial  magnetic field strength and
density of the shell; both measured in the lab frame. Note that, since 
the shell is assumed initially cold, we neglect the contribution of 
the internal energy to the kinetic flux of the shell. 

The total energy of the shell is
\be
E=4\pi R_0^2\Delta_0 (\gamma_0\rho_0
c^2+B_0^2/4\pi)=E_k(1+\sigma_0),  
\label{energy}
\ee
where the first term stands for the kinetic energy and the second for the
magnetic energy of the shell. In other words, the magnetization $\sigma_0$
parameterizes the fraction of the total energy of the flow that is
in kinetic and in magnetic form. These fractions are  $1/(1+\sigma_0)$ and
$\sigma_0/(1+\sigma_0)$ respectively.

Once $E$, $\Delta_0$, $\gamma_0$ and $\sigma_0$ are fixed the
properties of the ejecta are defined. The reference values that
we adopt for these parameters are summarized in the
following.  In our spherically symmetric model $E$ is roughly equal
to the isotropic equivalent energy $E_{\rm iso}$ of the burst and,
thus $E\sim E_{\rm iso} \sim 10^{53}E_{53}\,$erg (with spread of a
couple of orders of magnitude from burst to burst). The width of the
shell can be related to the observed duration $T_{GRB}$ of the
GRB through the relation $\Delta_0=c T_{GRB}\sim
10^{12}\Delta_{12}\,$cm for a $\sim 30\,$s burst.  This connection
between observed burst duration and shell width holds in the internal
shock model (Rees \& M\'esz\'aros 1994; Nakar \& Piran 2002) and
photospheric models (as discussed, for example, in Giannios \& Spruit
2007) for the prompt emission. The Lorentz factor of the flow is less
constrained. From theoretical arguments related to the ``compactness
problem'', it is expected to be $\gamma_0\simmore 100$. We adopt 
  $\gamma_0=10^{2.5}\gamma_{2.5}$ as reference value.  Among the 
shell parameters, $\sigma_0$ is the most
model dependent quantity and we deal with it as a free parameter.
In fireball models $\sigma_0\ll 1$ while in PDFs $\sigma_0\simmore 1$. 
                                    
Before studying the initial interaction of arbitrarily magnetized ejecta with
the external medium, we summarize the results of the unmagnetized $\sigma_0=0$
case. These results facilitate the presentation of the more general case where
$\sigma_0$ is arbitrary.

\subsubsection{Non-magnetized ejecta}

The deceleration of {\it non-magnetized} ejecta has been well studied 
with both analytical  (Sari \& Piran 1995; M\'esz\'aros \& Rees 1997a) 
and numerical approaches with 1-dimensional (Kobayashi et
al. 1999) and 2-dimensional (Granot et al. 2001; Cannizzo et al. 2004; Meliani et al. 2007)
hydrodynamical simulations. The 2-dimensional studies are important to follow 
the late-time lateral spreading of collimated ejecta. Here, 
we are interested in the initial phases of the deceleration of the ejecta.
These can be followed assuming spherical symmetry since the initial Lorentz 
factor of the flow is likely larger than the inverse of the jet opening angle
and, therefore, lateral spreading is unimportant.    

The hydrodynamical evolution of the shell-external medium
configuration is rather well understood. The interaction of two media
leads to a forward shock that propagates into the external medium. At
the same time a second shock (the so called reverse shock) propagates
into the ejecta. The formation of the reverse shock always takes place
in cold, non-magnetized ejecta since the shocked material moves
supersonically with respect to the unshocked shell. This can
be formally shown by solving the initial Riemann problem at the 
ejecta-external medium interface (see, for example
Rezzolla \& Zanotti 2001). We will see that this is not always the
case in strongly magnetized ejecta where MHD waves can transfer
information at relativistic speeds.  The shell can, thus, smoothly
adjust its properties upon interacting with the external medium.

The strength of the reverse shock depends on the ratio of the
densities of the shell and the external medium and on the bulk Lorentz
factor of the flow. It can be shown (Sari \& Piran 1995) that 
the strength of the reverse shock can be conveniently parameterized
by
\be \xi\equiv \sqrt{\frac{l}{\Delta_0}}\frac{1}{\gamma_0^{4/3}},
\label{ksi}
\ee 
where $l=(3E/4\pi n_{\rm e} m_{\rm p}c^2)^{1/3}$ is the Sedov length.  In the limit
  where $\xi\gg 1$ the reverse shock is Newtonian and, if $\xi\ll 1$,
  relativistic (Sari \& Piran 1995). 

The $\xi$ quantity can be related to the more familiar 
``deceleration radius'' $r_{\rm dec}$ and the ``spreading  
radius'' $r_{\rm s}$. The first
is defined as the radius where the ejecta accumulate from the external medium
a mass $\gamma_0^{-1}$ times their own mass, resulting in $r_{\rm
dec}=l/\gamma_0^{2/3}$. The $r_{\rm s}$ is the radius
where a shell with mildly relativistic proper motions starts 
spreading radially $r_{\rm s}=\Delta_0\gamma_0^2$ (see Sect. 3.1).
Using the last expressions, $\xi$ can be expressed as
\be 
\xi=\sqrt{\frac{r_{\rm dec}}{r_{\rm s}}}.
\ee
In terms of $r_{\rm dec}$ and $r_{\rm s}$, the condition for a relativistic 
reverse shock is that $r_{\rm dec}\ll r_{\rm s}$.

Using as reference  value for the external medium number density 
$n_e=10^0n_0$ cm$^{-3}$ we have that 
\be
\xi=0.73\frac{E_{53}^{1/6}}{n_0^{1/6}\Delta_{12}^{1/2}\gamma_{2.5}^{4/3}},
\ee 
showing that $\xi$ possesses a very weak dependence with $E_{53}$
  and $n_0$.  For various GRB parameters $\xi$ can typically vary in
the $0.1\simless \xi\simless 10$ range allowing for both Newtonian and
relativistic reverse shocks in different bursts.

In the Newtonian reverse shock limit, the ejecta do not 
decelerate much during the reverse shock crossing. They enter
the deceleration phase once they accumulate from the external medium mass
that is a factor $\gamma_o$ smaller than their own mass. 
In the relativistic reverse shock case
substantial deceleration takes place already when the reverse shock crosses the ejecta.    
In both relativistic and Newtonian reverse shock limits the radius
of the reverse shock crossing $r_{{\rm rs}^H}$ is given by the
expression\footnote{The reverse crossing radius differs in the two limits by
a small factor which we ignore here.} 
(Sari \& Piran 1995)
\be
r_{{\rm rs}}^H=l^{3/4}\Delta_0^{1/4}=(r_{\rm s}r_{\rm dec}^3)^{1/4},
\label{r_rs}
\ee
where the index ``H'' stands for hydrodynamical (non-magnetized) case.

In the Newtonian reverse shock case, the radius of the reverse shock
crossing is modified if one allows for mildly relativistic relative speeds
within the shell. The relative motions lead to shell spreading and
strengthening of the reverse shock. This effect is considered in Section 3.1 

After the reverse shock reaches the back part of  the ejecta, 
there is an initial phase of interaction where shocks and rarefaction 
waves cross the ejecta. Gradually most of the energy is passed in the shocked external medium
and the whole structure relaxes to the self-similar
structure described in Blandford \& McKee (1976).
From this point on, the evolution of the blastwave depends only on the 
energy $E$ and the density of the external medium $n_{\rm e}$.

\subsubsection{Arbitrarily magnetized ejecta}

The dynamics of the deceleration of magnetized ejecta has not been
studied in the same detail.  To understand the various phases of the
deceleration of the ejecta and the transfer of their energy into the
shocked external medium one needs to follow the dynamics of 
ejecta-external medium interaction in detail. The internal evolution of the
shell is particularly important in this respect. The study of
the complete deceleration dynamics is the goal of the relativistic MHD
simulations to be reported elsewhere (Mimica et al. 2007).  Here
we are concerned with the very initial phase of the interaction where
the reverse shock is (or is not) crossing the ejecta and derive the
conditions for such shock to form.

The initial phase of reverse shock crossing has been studied by Zhang \&
Kobayashi (2005) by solving the {\it ideal} MHD shock 
conditions for arbitrarily magnetized ejecta with a dominant toroidal field
(see also Fan et al. 2004 for the case of mildly magnetized ejecta).  
Their analysis describes the reverse shock crossing phase {\it provided that there is
  a reverse shock forming}. They show that the reverse shock crosses
faster ejecta of high magnetization $\sigma_0$ with respect to
un-magnetized ejecta of the same $E$, $\Delta_0$, $\Gamma_0$.
They show that the reverse shock crossing radius scales as $r_{{\rm
    rs}}\propto 1/\sqrt{\sigma_0}$ for $\sigma_0\gg 1$ and can be
written \be r_{{\rm rs}}=l^{3/4}\Delta_0^{1/4}C_{\Delta}=(r_{\rm s}r_{\rm
  dec}^3)^{1/4}C_{\Delta}.
\label{rsB}
\ee
The coefficient $C_\Delta$ depends mainly on the magnetization $\sigma_0$
and very weakly on the relative Lorentz factor of the shocked-unshocked
ejecta (see Fig.~3 in ZK05). For the purposes of this work, we adopt the expression 
$C_\Delta \sim 1/\sqrt{1+\sigma_0}$ which is rather
accurate in reproducing their results. 
Note that this expression for $C_\Delta$ gives also the correct 
zero magnetization limit (\ref{r_rs}). 

\section{Condition for existence of a reverse shock}

Cold, non-magnetized ejecta are always crossed by a reverse shock upon
interacting with the external medium. This is the case since the
sound speed of the ejecta is low and does not allow for fast transfer of the 
information of the interaction with the external medium throughout their volume. 
On the other hand, in a flow that is strongly
magnetized and sub-fast magnetosonic (as in the Lyutikov \& Blandford 2003
model) there is no reverse shock forming. The flow
adjusts gradually to the changes of the pressure in the contact
discontinuity that separates the magnetized flow from the shocked external medium. 
Here we generalize to arbitrarily magnetized ejecta and derive the condition
for the formation of a reverse shock.

An important radius to be considered for the existence of a reverse shock is the ``contact'' 
radius $r_{\rm c}$ at which fast MHD waves cross the width of the shell.
At this radius, the information on the conditions of the front part 
has been transfered to the rest of the shell.
After this radius is crossed the magnetized shell comes in pressure balance
and adjusts its structure on a timescale shorter than the expansion
time. Any gradual changes of the physical properties at the contact
discontinuity that separates the shell from the shocked external medium
lead to smooth evolution of the shell properties. The very existence
of the contact radius implies that different parts of the flow have dropped
out of MHD contact during the acceleration phase. Our analysis applies for
super-fast magnetosonic flows (i.e. flows with $\gamma_0>\sqrt{1+\sigma_0}$) 
for which this is the case. 

A second important radius for the existence of a reverse shock 
is $r_{{\rm rs}}$ (see eq.~(7)). If $r_{\rm c} >  r_{{\rm rs}}$, 
a reverse shock forms and crosses the ejecta before MHD contact 
within the shell is established.
In this case, internal evolution does not take place in the shell before the reverse
shock crossing. On the other hand, if $r_{\rm c}< r_{{\rm rs}}$, there is no reverse shock 
forming and $r_{\rm rs}$ does not have a physical meaning.
The condition for the formation of a reverse shock is, thus, to set
$r_{\rm c}>r_{{\rm rs}}$. Note that this condition states the obvious
fact that there is a reverse shock forming as long as it propagates
supersonically into the ejecta. As we discuss in Section 3.3, 
a different condition for the existence of a reverse shock adopted 
by ZK05 leads to incorrect estimates on the magnetization needed to suppress
the formation of a reverse shock.

The ``contact'' radius is given by the distance the shell travels
before a fast-magnetosonic wave crosses the shell.  Any news from the
front travels into the shell with speed
$\beta_A=\sqrt{\sigma_0/(1+\sigma_0)}$ (with a corresponding Lorentz
factor $\gamma_A=\sqrt{1+\sigma_0}$) in the frame comoving with the
shell. As seen in the central engine frame, this wave moves with
Lorentz factor $\gamma_w$ given by the expression $\gamma_A\approx
(\gamma_w/\gamma_0+ \gamma_0/\gamma_w)/2$.  The last expression is
accurate for $\gamma_0$, $\gamma_w\gg 1$
and can be solved for $\gamma_w$ resulting in
$\gamma_w=\gamma_0(\sqrt{1+\sigma_0}-\sqrt{\sigma_0})$.  The shell
comes into contact when the wave crosses it. This takes place at
radius
\be 
r_{\rm c}=\frac{\Delta_0 c}{v_0-v_w}\simeq
\Delta_0\gamma_0^2\Big(\sqrt{\frac{1+\sigma_0}{\sigma_0}}-1\Big),
\label{rc}
\ee 
where we have used that $v_w/c\simeq 1-1/2\gamma_w^2$ and $v_0/c\simeq 1-1/2\gamma_0^2$.

The contact radius depends only on the shell properties and not on those
  of the external medium. The reason is that the speed
  of propagation of fast-magnetosonic waves depends only on the
  magnetization of the shell as a function of the radius $\sigma(r)$
  which turns out to be equal to the initial value $\sigma_0$. To
  check such an assertion, one needs to consider that a 
  shell that expands spherically decreases both its density and its
  magnetic energy density as $r^{-2}$. Therefore,
  $\sigma(r)$ remains uniform in the shell and equal to its initial
  value (eq.~1) until, at least, the shell reaches $r_{\rm c}$.

From eq.~(\ref{rc}) we see that in the limit of $\sigma_0\gg 1$,
$r_c\simeq \Delta_0\gamma_0^2/2\sigma_0\propto \sigma_0^{-1}$ while
for $\sigma_0 \to 0$, $r_c\to \infty$. In the hydrodynamical
($\sigma_0=0$) limit $r_{\rm c}> r_{{\rm rs}}$ and a reverse shock
crosses the ejecta at $r_{{\rm rs}}$.  On the other hand, since
the contact radius scales with the magnetization
as $r_{\rm c}\propto \sigma_0^{-1}$, it is reduced faster than the
$r_{{\rm rs}}\propto \sigma_0^{-1/2}$ for increasing $\sigma_0$.
There is always a magnetization for which $r_{{\rm rs}}$ and $r_{\rm c}$ 
become equal and the reverse shock is suppressed.

The condition for the existence of the reverse shock is given
by setting $r_{\rm rs}<r_{\rm c}$ and using eqs. (\ref{rsB}), (\ref{rc}) 
and the definition (\ref{ksi}) of $\xi$. 
The line that separates the reverse shock from the non-reverse shock regime 
in the $\xi$-$\sigma_0$ parameter space is
\be
\Big(\frac{r_{\rm dec}}{r_{\rm s}}\Big)^{3/2}
=\xi^3=\frac{1+\sigma_0}{\sigma_0}(1+2\sigma_0-2\sqrt{\sigma_0(1+\sigma_0)}).
\label{ksi-sigma}
\ee
In the limit of small and large $\sigma_o$ the last expression simplifies
respectively to
\begin{eqnarray}
\xi&=&\frac{1}{\sigma_0^{1/3}} \quad\rm{for}\quad   \sigma_0\ll 1,
\nonumber\\
\label{Limits}\\
\xi&=&\frac{1}{(4\sigma_0)^{1/3}} \quad\rm{for}\quad   \sigma_0\gg 1. \nonumber
\end{eqnarray}

Although, the expression $\xi=(4\sigma_0)^{-1/3}$ is formally
derived for $\sigma_0\gg 1$, it is still a fair approximation for
$\sigma_0\simmore 0.1$ and will be used in the following discussion.
Solving the last expression for $\sigma_0$, 
and using the reference values of the GRB parameters we find that
a reverse shock forms when
\be
\sigma_0\simless 0.6 n_0^{1/2}\Delta_{12}^{3/2}\gamma_{2.5}^4E_{53}^{-1/2}.
\label{rscondition}
\ee

The curve defined by eq.~(\ref{ksi-sigma}) is shown in Fig.~1 with a dashed
line that turns into solid for $\sigma_0\simmore 0.3$. 
It separates the $\xi$-$\sigma_0$ parameter space into regions with
the upper right one characterized by no reverse shock. 
In the $\sigma_0\ll 1$ regime the $\xi=1$ dash-dotted line separates the Newtonian
reverse shock ($\xi>1$) regime from the $\xi<1$ relativistic reverse shock
one. Note that a small magnetization suffices to suppress a reverse shock   
for  large $\xi$ and vice versa. 

\begin{figure}
\resizebox{\hsize}{!}{\includegraphics[angle=270]{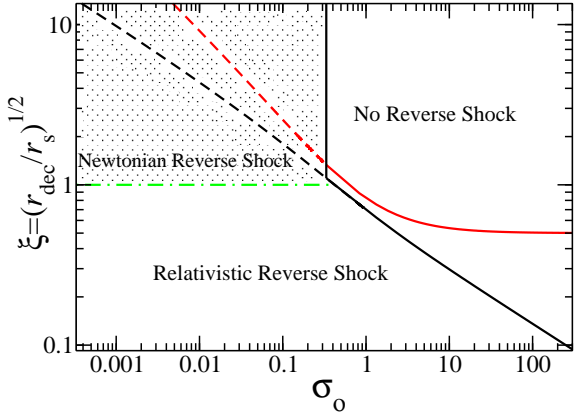}}
\caption[]{Regime of existence of a reverse shock in the
  $\xi$-$\sigma_0$ plane.  The $\xi$ is defined by eq.~(\ref{ksi})
  (or eq.~(4)) and $\sigma_0$ stands for the
  Poynting-to-kinetic flux ratio of the flow at the onset of the
  afterglow. Ignoring radial spreading of the ejecta, the lower left part with
  respect to the black, dashed line shows the regime where there is a reverse
  shock forming for a constant-density external medium. For $\sigma_0\simmore
  1$ there is no reverse shock forming for a large parameter space that 
describes the properties of the GRB ejecta. The solid line shows the frontier 
between ``no reverse shock'' and ``reverse shock'' when radial spreading of 
the shell is taken into account. In the case of shell spreading, the shaded area is characterized 
by a mildly relativistic reverse shock (see  Sect.~\ref{sec:shell-spreading}).
Similarly, the red lines show the frontier between ``no reverse shock'' and
``reverse shock'' in the wind-like external medium.}

\label{fig1}
\end{figure}

As it is obvious from eq.~(\ref{rscondition}),
for typical GRB parameters for a strongly magnetized flow with $\sigma_0\simmore 1$
there is no reverse shock forming. In this case, the onset of the afterglow emission will not be
characterized by an optical flash (M\'esz\'aros \& Rees 1997a; Sari \& Piran
1999) that is expected to be the signature of the reverse shock emission.
 Note, however that this conclusion depends sensitively on some of the GRB 
parameters. This is especially true for the Lorentz factor of the
GRB flow which is also, perhaps, the least constrained parameter. 
Had we used a reference value of $\gamma_0=1000$ for it then the transition
to no reverse shock would take place for $\sigma_0\sim$ a few tens.
This means that one can still have a reverse shock forming in
a $\gamma_0\simmore 1000$, $\sigma_0\gg 1$ flow. However, 
when a reverse shock forms in a high $\sigma_0$ flow it is
not expected to result in bright emission (at least in the ideal MHD
limit discussed here) since the shock dissipates only part
of the kinetic energy which is a small fraction of the total
energy of the shell (ZK05).       

So far, we have focused in ejecta with initially uniform motion, i.e. with no
radial spreading. Shell spreading introduces new features to the picture
presented so far. These features are the topic of the next section.

\subsection{Shell-spreading effects}
\label{sec:shell-spreading}
 
It is quite likely that the ejecta are characterized by relative
motions after the acceleration, collimation and internal interaction phases. 
These relative motions may well be mildly relativistic in the shell rest
frame leading to ejecta with moderate variations of the
Lorentz factor $\delta \gamma_0 \sim \gamma_0$ along their width.

Ejecta with their front part moving faster than their rear one with  moderate
variation of the bulk Lorentz factor start spreading at radius $r_s\simeq
\gamma_0^2 \Delta_0$ (Sari \& Piran 1995; Kobayashi et al. 1999). 
Because of the spreading, the width of the shell increases linearly 
with distance at larger radii. 
It has to be noted that $\Delta_0 / c$ does not necessary correspond to
the time during which the central engine is actively releasing energy
into the outflow. As pointed out by Aloy, Janka \& M\"uller (2005), in
case of short GRBs, the initial shell width $\Delta_e = c T_a$ ($T_a$
being the time during which the central engine actively releases
energy) may stretch radially by a factor $\sim 2-20$ due to the same
mechanism invoked for shell spreading at later times, namely, the
relative velocity difference between the forward and rear radial edges
of the ejecta.

In the case of non-magnetized ejecta with $\xi > 1$, the shell starts
spreading before the reverse shock crossing. Radial expansion leads to
density decrease in the shell which is faster than $1/r^2$. As a
result, by the end of the reverse shock crossing the shock is mildly
relativistic.  When $\xi < 1$, the reverse shock crosses the shell
before it reaches the spreading radius $r_{\rm s}$.
The relativistic reverse shock case is, hence, not affected by
the spreading of the shell since, by the time the shock has crossed
the shell the density of the latter has not dropped appreciably in 
comparison to the non-spreading case (Sari \& Piran 1995).
One may also express the condition of formation of a
relativistic reverse shock in terms of the deceleration radius
$r_{\rm dec}$ and $r_{\rm s}$, which reads $r_{\rm dec}\ll r_{\rm s}$.

Spreading of the shell influences the interaction of the ejecta
with the external medium for magnetized ejecta as well. If the contact
radius $r_{\rm c}$ is larger than the spreading radius, the shell spreads
before fast MHD waves produced in the front part of the shell catch up with the rear part. 
Using the estimate $r_s=\gamma_0^2 \Delta_0$ for the spreading radius and
eq.~(\ref{rc}) for $r_{\rm c}$, one finds that $r_{\rm s}\simless r_{\rm c}$ for
$\sigma_0\simless 0.3$. For low enough $\sigma_0$ the shell does not
come into MHD contact because of radial spreading. 
Under these conditions, a mildly relativistic RS forms just like in the $\sigma_0=0$ case. 

The shaded region of Fig.~1 shows the regime where spreading   
of the shell leads to the formation of a reverse shock. Quite generally
for $\sigma_0\simless 0.3$, there is always a reverse shock forming
in a shell that spreads because of mildly relativistic proper motions.

\subsection{Deceleration in a wind environment}

A possible environment of the deceleration of the GRB flow is that of stellar
wind (i.e. ``collapsar'' model for long GRBs; Woosley 1993).
The density profile of the external medium in this case is $n_{\rm w}=A/r^2$.
Here, we derive briefly the condition for the existence of a reverse
shock in such density profile. 

The reverse-shock crossing radius for un-magnetized ejecta that decelerate
in wind density profile is $r^{\rm H}_{\rm rs,w}\sim (r_{\rm s} 
r_{\rm dec,w})^{1/2}$ (Sari \& Piran 1995),
where the deceleration radius is defined as $r_{\rm dec,w}=l_{\rm w}/\gamma_0^2=E/4\pi Am_{\rm p}
\gamma_0^2c^2$. The $l_{\rm w}$ is the Sedov length for the wind density
profile. If the ejecta is magnetized, the reverse shock crosses
them faster with respect to the hydrodynamical case. In the
high-$\sigma_0$ limit the (reverse) shock conditions lead to
$r_{\rm rs,w}\simeq r^{\rm H}_{\rm rs,w}/\sigma_0$ (e.g. ZK05).
An approximate expression for the radius of the crossing of the reverse shock  
that reproduces both the $\sigma_0\to 0$ and $\sigma_0\gg 1$ limits
is
\be
r_{\rm rs,w}\sim \frac{(r_{\rm s} r_{\rm dec,w})^{1/2}}{1+\sigma_0}.
\label{rrsw}
\ee
The condition for existence of a reverse shock is that the
$r_{\rm rs,w}<r_{\rm c}$. The contact radius depends only on the shell
properties and not on the density profile of the external medium; 
it is defined by eq.~(\ref{rc}).
The line that separates the ``no reverse shock'' from the ``reverse
shock'' regime is found by setting $r_{\rm rs,w}=r_{\rm c}$. It results in:
\be
\xi_{\rm w}= \Big(\frac{ r_{\rm dec,w}}{r_{\rm s}}\Big)^{1/2}=(1+\sigma_0)
\Big(\sqrt{\frac{1+\sigma_0}{\sigma_0}}-1\Big).
\ee
The line defined by eq.~(13) is shown in the $\xi$-$\sigma_0$ plane of Fig.~1. 

If the shell is characterized by mildly relativistic proper motions, its
spreading radius is $r_{\rm s}\sim \Delta_0\gamma_0^2$. Just as in the 
case of constant density external medium, spreading of the shell
leads to the formation of a reverse shock when  $\sigma_0\simless 0.3$. 

\subsection{Non-ideal MHD effects}

In this work, we focus on the existence of  a reverse shock 
(and the associated emission) coming from the interaction of a 
shell, permitted by a large scale magnetic field, with the external medium. 
Furthermore, this analysis is based on the ideal MHD shock conditions 
derived by ZK05. However, non-ideal MHD effects may also be of relevance 
in the flow. 
   
It is possible, for example,  that the magnetized flow is launched with magnetic
fields that contain a random component; one that changes polarity on short scales in the 
radial direction (Thompson 1994; Drenkhahn \& Spruit 2002; Thompson 2006).  Furthermore
small scale field reversals may develop during the acceleration phases as
result of MHD instabilities in the jet (Giannios \& Spruit 2006). 
Unless magnetic reconnection is efficient in dissipating the random
component of the magnetic field (e.g. Drenkhahn \& Spruit 2002), small scale reversals 
of the field along the radial direction are still present in the afterglow
regime. Such field structure can have an effect on the shock conditions
and therefore on the condition for the existence of a reverse shock 
(for discussion see Thompson 2006).

Dissipation of part of the magnetic energy that is contained in the
flow at large radii can result to particle acceleration and emission
with a variety of observational consequences. This emission is {\it
  not} directly related to shock-accelerated particles. The reverse
shock can, for example, lead to forced reconnection of a reversing
magnetic field. The released energy can power the prompt GRB emission
(Thompson 2006). Magnetic dissipation can also take place during a
later stage of the deceleration of the ejecta. In this picture,
slowing down of the ejecta leads to the revival of MHD instabilities
in regions of the flow that come into causal contact again. Such
delayed dissipation can power the afterglow flares (Giannios 2006a)
observed in many early X-ray afterglows (Burrows et al. 2005; Zhang et
al. 2006).

\subsection{Comparison with previous work}

The early phases of interaction of the GRB ejecta have been previously
studied in the hydrodynamical limit (i.e. $\sigma_0\to 0$; Sari \& Piran 1995)
and for a sub-fast magnetosonic, strongly magnetized, flow 
(i.e. $\gamma_0^2<1+\sigma_0$; Lyutikov \& Blandford
2003; Genet et al. 2006). The outcome of those studies is that in the former
limit there is always a reverse shock developing, in the latter the flow
adjusts smoothly to the properties of the shocked external medium and there is
no reverse shock. Those results are in agreement with these reported here.

The intermediate range of magnetization has been studied in ZK05
by solving for the shock conditions for the forward and the
reverse shock. Their study is applicable for the parameter space
where there is a reverse shock forming. They also estimate the conditions
for the formation of a reverse shock assuming a constant density external 
medium. They find that large values of $\sigma_0\sim
100$ are needed (for the reference GRB parameters) to suppress the existence of a
reverse shock. This estimate is more
than two orders of magnitude different from the one
presented here. To probe the difference, we briefly sketch the ZK05 
condition for the existence of a reverse shock. 
 
ZK05 define the deceleration radius as the radius where the ejecta
accumulates from the external medium mass a factor $1/\gamma_0$ less
than the external medium mass:
$r_d^{ZK}=l/\gamma_0^{2/3}(1+\sigma_0)^{1/3}$.  The conditions for the
existence of a reverse shock that they use are that (i) the flow speed
exceeds the fast magnetosonic speed (i.e.
$\gamma_0>\sqrt{1+\sigma_0}$) and (ii) the magnetic pressure of the
shell is lower than that of the shocked external medium at the
deceleration radius $r_d^{ZK}$: i.e. that $B_0^2/8\pi
\gamma_0^2<(4/3)\gamma_0^2 \rho_e c^2$.  In the last expression, one
uses that the comoving magnetic field is $B_{0,co}=B_0/\gamma_0$ and
the forward relativistic shock condition that gives for the gas
pressure of the shocked external medium: $P_{\rm g}=(4/3) \gamma_0^2\rho_{\rm e}
c^2$.  Using eq.~(2) and focusing in the $\sigma_0 \gg 1$ limit, the
condition for $\sigma_0$ for the existence of a reverse shock can be
written $\sigma_0 \simless 190/\xi^3$. The boundary that separates the
shock from the non-shock regime is $\sigma_{ZK}\simeq 190/\xi^3$. It
is very different with respect to that of eq.~(\ref{Limits}) derived
here.

What is the source of this discrepancy? ZK05 apply their condition in
the deceleration radius and not the reverse shock crossing radius
which we believe is the relevant one so as to check the existence of a
reverse shock\footnote{Note also that the definition of the
deceleration radius is controversial (see Lyutikov's 2005
comments)}.  Furthermore, the condition itself for the existence of
the reverse shock used in ZK05 is questionable. For
magnetization $\sigma_0\sim \sigma_{ZK}$ the ``contact'' radius $r_{\rm c}$
is much shorter than the ``deceleration'' radius $r_d^{ZK}$ and the
reverse shock crossing radius $r_{\rm rs}$. This means that one gets the
erroneous result that the reverse shock moves subsonically with
respect to the unshocked material. The fact that $r_{\rm c}<r_d^{ZK}$
indicates instead that much before the flow reaches $r_d^{ZK}$ 
the shell has already evolved. The spherical geometry starts to play an
important role and the problem becomes that of dynamical
evolution of the shell properties and not just shock conditions of a
non evolving shell. 

On the other hand, when a reverse shock does form then it crosses
the shell before its different parts have time to come into contact through
exchange of MHD waves. This means that the properties of the shell
upstream to the shock do not evolve much before the shock crossing.
In this case the analysis of ZK05 is applicable. This justifies the 
use of their result on the  reverse shock crossing radius given by eq.~(\ref{rsB}).

\section{Discussion}

The role of magnetic fields in GRB flows remains uncertain.
In fireball models, the GRB flow is launched with dynamically unimportant
magnetic fields while MHD energy extraction can lead to a 
Poynting-flux dominated flow. Observations of the prompt emission alone appear 
hard to distinguish between fireball or PDF models\footnote {with the possible exception
of multi GeV prompt emission that may show signature of 
pair cascades in a neutron rich fireball that is absent in PDF flows; 
Koers \& Giannios (2007).}. Early afterglow observations can, however, probe 
the magnetic content of the ejecta. The magnetization can leave its signature in the
early forward shock (Lyutikov \& Blandford 2003; Genet et al. 2006)   
and the reverse shock emission (Kumar \& Panaitescu 2003; Zhang et al. 2003; 
Fan et al. 2004; ZK05). 

\subsection{Magnetization and existence of a reverse shock}

Here, we derive the condition for the existence of a reverse shock in
the ejecta that contain a large scale magnetic field, assuming the
ideal MHD description is adequate. The ejecta decelerate interacting
with the circumburst medium. External medium with both constant 
and wind-like density profiles are considered. It is shown that, 
for high enough magnetization, the ejecta interact smoothly with the
external medium and decelerate without the formation of a reverse shock. The
conditions for the formation of a reverse shock are derived by
demanding that the reverse shock propagates supersonically in the
ejecta. The conditions are summarized by eqs.~(\ref{ksi-sigma}) and
(\ref{rscondition}) and Fig.~1. Our results are
substantially different from those of ZK05 who do not consider the
internal evolution of the ejecta when deriving their criterion for the existence of a
reverse shock.

One important finding of the study is that $\sigma_0 \simmore 1$ suffices
to suppress the formation of a reverse shock for a large parameter space
that characterizes GRB flows. This parameter space is larger for
a constant density external medium in comparison to the wind case.
This means that only a small fraction of PDFs are expected to show reverse 
shock emission. Even when a reverse shock forms in a high $\sigma_0$
 flow, it is not expected to result in bright emission (at least in the ideal MHD
limit discussed here) since the shock dissipates only part
of the kinetic energy which is a small fraction of the total
energy of the shell (see, e.g., ZK05). On the other hand, allowing for 
radial spreading of the ejecta, practically all the flows with 
$\sigma_0 \simless 0.3$ are expected to be crossed by a mildly-relativistic 
or relativistic reverse shock.
    
This work contributes in understanding the early deceleration dynamics
of magnetized ejecta. The complete evolution of the system is not considered
here. Important questions related to the timescale over which the magnetic
energy of the shell (which may dominate the total energy budget) is passed 
into the shocked external medium remain unexplored. We are currently attacking
these questions through relativistic MHD simulations to be reported elsewhere 
(Mimica et al. 2007).

\subsection{Connection to observations}
   
A clear prediction of the fireball models is that there is a reverse
shock crossing the ejecta on the onset of the interaction of the
ejecta with the external medium. The reverse shock emission is
expected to result in a short lived, bright emission
event. Although optical flashes have been observed in a few bursts
(e.g. GRB990123, Akerlof et al. 1999; GRB021211, Fox et al. 2003) and
are believed to be powered by the reverse shock emission, they are
rare events. This may be a result of a more luminous forward shock
masking the reverse shock optical emission (McMahon et al. 2006;  
Mundell et al. 2007b).

Our analysis points to a different possibility, i.e. that the paucity of 
bursts with clear  reverse shock emission is a result of the frequent reverse
shock suppression. It is shown here that this happens for a large GRB
parameter space in a strongly magnetized flow. 
Moreover, in cases where reverse shock emission is 
identified, modeling of the microphysical parameter connected to the
magnetic field strength indicates ejecta with dynamically important
magnetic fields (Fan et al. 2002; Kumar \& Panaitescu 2003; 
Zhang et al. 2003; ZK05) in agreement with our suggestion.
  
Polarized emission, result of synchrotron emitting, (reverse)
shock-accelerated particles is expected 
from a coherent magnetic field. Note, however, that the 
lack of significant optical polarization in the early afterglow does not
necessarily show the absence of large scale field in the ejecta but may be
result of the absence of reverse shock (suppressed by the strong field). 
This is particularity likely if there is no evidence for the distinct emission 
patterns expected from a reverse shock.
In light of these considerations the claim of Mundell et al. (2007a) that
the low early polarization signal in GRB060418 rules out the presence of large
scale magnetic field in the ejecta is premature. 

This study has focused in the ideal MHD limit. On the other hand
non-ideal MHD effects can lead to additional energy release and emission during the 
afterglow phases of strongly magnetized flows.
This emission is {\it not} directly related to shock accelerated particles 
and therefore to the existence or not 
of a reverse shock. On the other hand, the crossing of the reverse shock
can lead to forced reconnection in the ejecta that powers the GRB emission
(Thompson 2006). Furthermore, the deceleration of the ejecta can lead to 
revival of MHD instabilities that lead to magnetic dissipation. Such 
late-time energy release can power the afterglow flares observed in 
many early X-ray afterglows (Giannios 2006a). 

\begin{acknowledgements}
DG thanks Henk Spruit for valuable discussions.
PM is  at the University of Valencia with a 
European Union Marie Curie Incoming International Fellowship (MEIF-CT-2005-021603). 
MAA is a Ram\'on y Cajal Fellow of the Spanish Ministry of
Education and Science. He also acknowledges partial support from
the Spanish Ministry of Education and Science
(AYA2004-08067-C03-C01).

\end{acknowledgements}

\end{document}